\documentclass[onecolumn,prd,showpacs,preprintnumbers,amsmath,amssymb,floatfix]{revtex4}

\usepackage{graphicx}

\usepackage{bm}
\usepackage{amsfonts}
\usepackage{array}
\usepackage{microtype}
\usepackage{float}
\usepackage[normalem]{ulem}
\usepackage{epstopdf}
\usepackage{mathrsfs}
\usepackage{color}

\usepackage{xcolor}

\usepackage{pifont}

\usepackage{lineno,hyperref}

\hypersetup{colorlinks=true, linkcolor=red, filecolor=magenta, urlcolor=blue, citecolor=blue}

\begin{document}

\title{A relativistic compact stellar model of anisotropic quark matter mixed  
with dark energy}

\author{Theophanes Grammenos}
 \email{thgramme@civ.uth.gr}
\affiliation{Department of Civil Engineering, University of Thessaly, Volos, Greece}

\author{Farook Rahaman}
\email{rahaman@associates.iucaa.in}
\affiliation{Department of Mathematics, Jadavpur University, Kolkata 700092, West Bengal, India}
	
\author{Saibal Ray}
\email{saibal@associates.iucaa.in}
\affiliation{Department of Physics, Government College of Engineering and
	Ceramic Technology, Kolkata 700010, West Bengal, India}

\author{Debabrata Deb}
\email{debabratadeb@iisc.ac.in}
\affiliation{Department of Physics, Indian Institute of Science, Bangalore 560012, India}

\author{Sourav Roy Chowdhury}
\email{roic@sfedu.ru}
\affiliation{Institute of Physics, Southern Federal University, 194 Stachki, Rostov-on-Don 344090, Russia}

\date{\today}

\begin{abstract}
The possibility of strange stars mixed with dark energy to be one of the candidates for dark energy stars is the main issue of the present study. Our investigation shows that quark matter acts as dark energy after a certain yet unknown critical condition inside the quark stars. 
Our proposed model reveals that strange stars mixed with dark energy feature a physically acceptable stable model mimic characteristics of dark energy stars. The plausible connections are shown through the mass-radius relation as well as the entropy and temperature. We particularly note that a two-fluid distribution is a major reason for the anisotropic nature of the spherical stellar system.  \\

\end{abstract}


\maketitle

\section{Introduction}
The study of the different aspects of quark matter has drawn attention among the astrophysicists and particle physicists in the last two decades. Bhattacharyya and his coworkers~\cite{Bhattacharyya1999,Bhattacharyya2000,Bhattacharyya2003} proposed that after the few microseconds of the big bang the universe undergoes a quark-gluon phase transition which may be a process of origin and survival of the quark matter. The nature of the confining force which triggered this phase transition is almost unknown. Witten~\cite{Witten1984} first considered that at the critical temperature $T_c \simeq 200~MeV$ a small portion of the colored objects like quark and gluon leaves hadronization through a phase transition to form the colored particles called quark nuggets (QN). These QNs are made of $u$, $d$ and $s$ quarks and have a density that is a few times higher than the normal nuclear density. This was further investigated by~\cite{Applegate1985,Farhi1984,Chandra2000}. For both cases of the compact stars and the early universe, Ghosh~\cite{Ghosh2008} investigated the role of quark matter in the phase transition. It is believed that quark matter exists in the core of neutron stars~\cite{Perez-Garcia2010}, in strange stars~\cite{Drake2002} and as small pieces of strange matter~\cite{Madsen1999}. Investigations by Rahaman et al.~\cite{Farook2012} and Brilenkov~\cite{Brilenkov2013}  led to an interesting and important result according to which quark matter plays the same role as dark energy on the global level. It is worth mentioning that for the last several years  LHC at CERN has been trying to recreate the situation encountered before and early in the hadronization period, performing collisions of the relativistic nuclei~\cite{Adamovich1991}. 

Chapline~\cite{Chapline2004} proposed that a gravitationally collapsing compact star with a mass greater than a few solar masses, a quantum critical surface for space-time and an interior region consisting of a large amount of dark energy compared to the ordinary space-time can be defined as a dark energy star. He also predicted that this surface of a compact dark energy star is a quantum critical shell~\cite{Chapline2001}. When ordinary matter which has energy beyond the critical energy $Q_0$ enters the quantum critical region, it decays into constituent products and corresponding radiation is emitted in the outward direction perpendicular to that quantum critical surface. For the matter having energy less than $Q_0$ these constituent products and the radiation can pass through that critical surface and follow a diverging geodesic inside the star. For compact objects and compact stars at the center of galaxies the energy of the quarks and gluons inside the nucleons is higher than the critical energy $Q_0$~\cite{Barbierii2004}. According to the Georgi-Glashow grand unified model nucleons decay in a process in which a quark decays into a positron and two antiquarks. So the observation of the excess positron in the center of the galaxy may validate the presence of dark energy stars. 

In the present article we have tried to investigate the possible connection between the proposed quark star model mixed with dark energy and dark energy stars. Following the works of  Rahaman et al.~\cite{Farook2012} and Brilenkov~\cite{Brilenkov2013} we propose that quark matter is one of the possible candidates for dark energy. We are considering an anisotropic quark star model where we assume that the dark energy density is linearly proportional to the quark matter density. The proposed stellar configuration consists of two kinds of  fluid: (i) quark nuggets (QN), and (ii) dark energy having a repulsive nature. We have avoided any interaction between the fluids for the sake of the simplicity of the model. To describe the equation of state (EOS) for the effective fluid of the stellar model we have used the MIT bag EOS. At this point, a short comment on the mechanism that allows for the accumulated dark energy inside the star is needed. A possibility would be that a mechanism similar to the one responsible for dark matter accretion might be active. In such a case, a possible candidate would be weakly interactive massive particles (WIMP) that can accrue. At the same time, they have their antiparticles so that they can annihilate to create a heat source. Since dark energy density is expected to be very high, this heating would control the internal structure of the stars. However, one should keep in mind that WIMP’s interact weakly with baryons, whereas a dark energy particle most likely exhibits only gravitational interactions with baryons. Therefore, eventually, the accretion rate, as mentioned earlier, would be much lower than expected. In any case, all this is a theoretical conjecture. In reality, the mechanism that allows for the accumulated dark energy inside the stars remains unknown, which is also beyond the scope of the present work to discuss and needs further investigation.

It is worth mentioning that the anisotropy in compact stars may arise due to phase transition, the mixture of two fluids, the
existence of type 3A superfluid, bosonic composition, rotation, pion condensation etc., at the microscopic level. In the present study of the proposed anisotropic two-fluid model we consider the compact stars $PSR~J1614-2230$, $Vela~X-1$, $PSR~J1903+327$, $Cen~X-3$ and $SMC~X-4$ as testing candidates.

\section{Field equations}
To describe the interior of a relativistic compact star mixed with dark energy, we are considering
the following space-time line element
\begin{equation}
{ds}^{2}={{e}^{\nu(r)}}{{dt}^{2}}-{{e}^{\lambda(r)}}{{dr}^{2}}-{r}^{2}({{d\theta}^{2}}+{{\sin}^{2}}\theta{{d\phi}^{2}}),
\label{eq2}
\end{equation}
where $\nu$ and $\lambda$ depend only on the radial coordinate $r$.

The energy-momentum tensor components of the proposed two fluid model are given by
\begin{eqnarray}\label{eq03}
T^{\mu\nu}_m = (\rho^\textit{{eff}}+{p^\textit{{eff}}_t})u^\mu u^\nu-{p^\textit{{eff}}_t}g^{\mu\nu}+\left({p^\textit{{eff}}_r}-{p^\textit{{eff}}_t}\right)v^\mu v^\nu,
\end{eqnarray}
where
\begin{eqnarray} \label{eq003}
& {{\rho}^{eff}}= {{\rho}^{Q}}+{{\rho}^{DE}}, \\ \label{eq013}
& {{p_r}^{eff}} = ({{p}^Q_{r}}+{{p}^{DE}}), \\ \label{eq3}
& {{p_t}^{eff}} = ({{p}^Q_{t}}+{{p}^{DE}}).
\end{eqnarray}
${{\rho}^{Q}}$, ${{p}^{Q}_{r}}$ and ${p^Q_{t}}$ represent the quark matter density, the radial pressure and the tangential pressure respectively, whereas ${\rho}^{DE}$ and ${p}^{DE}$ represent, respectively, the dark energy density and the radial pressure within the star. On the other hand, by ${\rho}^{eff}$, ${p_r}^{eff}$ and ${p_t}^{eff}$ we represent, respectively, the effective energy density, the effective radial pressure and the effective tangential pressure of the matter distribution of the stellar system. 

Using Eqs.~\eqref{eq2} and~\eqref{eq03} the Einstein field equations for the present spherically symmetric anisotropic compact star read:
\begin{eqnarray}\label{41}
{{\rm e}^{-\lambda}} \left( {\frac {{\lambda}^{\prime}}{r}}-\frac{1}{r^{2}} \right) +{\frac{1}{{r}^{2}}}=8\,\pi\,
{{\rho}^{eff}},\\ \label{42}
{{\rm e}^{-\lambda}}\left({\frac {1}{r^{2}}}+\frac{{\nu}^{\prime}}{r} \right)-{\frac{1}{{r}^{2}}}=8\,\pi\,{p_r^{eff}},\\ \label{43}
\frac{1}{2}{{\rm e}^{-\lambda}}\left[\frac{1}{2}{\left({\nu}^{\prime}\right)}^{2}+{\nu}^{\prime\prime}-
\frac{1}{2}{\lambda}^{\prime}{\nu}^{\prime}+\frac{1}{r}\left({\nu}^{\prime}-{\lambda}^{\prime}\right)\right]=8\,\pi\,{p_t^{eff}}.
\end{eqnarray}

Now, to solve the Einstein equations for our spherical distribution we consider the following {\it ans{\"a}tze}: (i) the strange quark matter (SQM) distribution obeys the equation of state (EOS) of the phenomenological MIT bag model, i.e. 
\begin{eqnarray}\label{EOS1}
p^Q_r = \frac{1}{3}({\rho}^Q-4\,B_{g}),
\end{eqnarray}
where $B_{g}$ is the bag constant, and (ii) the dark energy radial pressure is related to the dark energy density as 
\begin{eqnarray}\label{EOS2}
{p}^{DE}=-{\rho}^{DE}.
\end{eqnarray}

Here the second \textit{ansatz} represents the EOS of the matter distribution, which is called  `degenerate vacuum' or `false vacuum'~\cite{Davies1984,Blome1984,Hogan1984,Kaiser1984}.

To close the system of equations following Mak and Harko~\cite{Harko2002} we consider a functional form for $\rho^q$ and $\rho^{DE}$ as follows
\begin{eqnarray}\label{44}
& \rho^Q=\rho_c\left[1-\left(1-\frac{\rho_0}{\rho_c}\right)\frac{r^2}{R^2}\right],\\ \label{45}
&\hspace*{-1.1cm} \rho^{DE} = \alpha \rho_c \left(1-\frac{r^2}{R^2}\right),
\end{eqnarray}
where both the functional forms for $\rho^Q$ and $\rho^{DE}$ are considered in such a way that they have a maximum value at the stellar core and decrease gradually to attain their minimum values at the surface. Note that the density for strange quark matter (SQM) at the surface has a nonzero finite value, i.e., $\rho_0\neq 0$, which has been taken care of in the assumed functional form of $\rho^Q$ (see Eq. (\ref{44})). $\rho_c$ denotes the central density of SQM, and $\alpha$ controls the amount of DE matter distribution corresponding to SQM, which also plays an important role in determining phase transition from the quark matter to the dark energy. $R$ denotes the total radius of the stellar object.

Now, from Eq.~(\ref{41}) we have
\begin{equation}
{\rm e}^{-\lambda}=1-\frac{2m(r)}{r},\label{37}
\end{equation}
where $m(r)$ is the mass function of the star, which is defined as follows:
\begin{equation}
m \left( r \right) =4\,\pi\,\int_{0}^{r}\!{{\rho}^{eff}} \left( r \right) {r}^
{2}{dr}.\label{38}
\end{equation}

The effective gravitational mass of the star can be found using Eqs. (\ref{44}-\ref{45}) and (\ref{38}) as
\begin{equation}
M=\frac{4}{15}\pi \,R \left( 2\,{R}^{2}\alpha\,\rho_{{c}}+3\,{R}^{2}\rho_{{0}}+2\,{R}^{2}\rho_{{c}} \right) . \label{eq16}
\end{equation}

Taking a vanishing radial pressure at the surface we find from the \textit{ansatz} (i), (ii) and  Eqs. (\ref{eq013}), (\ref{44}) and (\ref{45}) 
\begin{equation}
\rho_{{0}}= {4\, B_{{g}}}. \label{eq17}
\end{equation}

Using Eqs.~(\ref{41})-(\ref{43}) the complete set of structure equations is given by
\begin{eqnarray}
&\hspace*{-4.5cm} \frac{\rm{d}m}{\rm{d}r}=4\pi r^2 {{\rho}^{\textit{eff}}},\\ 
& -\frac{\rm{d}p^{\textit{eff}}_r}{\rm{d}r}-\frac{1}{2}\left(\rho^{\textit{eff}}+p^{\textit{eff}}_r\right)\nu^\prime+\frac{2}{r}\left(p^{\textit{eff}}_t-p^{\textit{eff}}_r\right)=0.
\end{eqnarray}

Now, substituting Eqs.~\eqref{42}, \eqref{EOS1}-\eqref{37} into \eqref{43} one can easily get an expression for $p^{eff}_t$ which yields the expected form of anisotropy ($\Delta=p^{eff}_t-p^{eff}_r$) as follows
\begin{eqnarray}
& \Delta(r)=-{\frac {{r}^{2}}{3\,{R}^{2}}}\big[-120\,\pi \,{R}^{2}{\alpha}^{2}{r}^{2}{\rho_{{c}}}^{2}+72\,\pi \,{\alpha}^{2}{r}^{4}{\rho_{{c}}}^{2}+80\,{\it B_g}\,\pi \,{R}^{4}\alpha\,\rho_{{c}}-96\,{\it B_g}\,\pi \,{R}^{2}\alpha\,{r}^{2}\rho_{{c}}\nonumber \\
& -80\,\pi \,{R}^{4}\alpha\,{\rho_{{c}}}^{2}-120\,\pi \,{R}^{2}\alpha\,{r}^{2}\rho_{{0}}\rho_{{c}}+96\,\pi \,{R}^{2}\alpha\,{r}^{2}{\rho_{{c}}}^{2} +48\,\pi \,\alpha\,{r}^{4}\rho_{{0}}\rho_{{c}}\nonumber \\
&-48\,\pi \,\alpha\,{r}^{4}{\rho_{{c}}}^{2}+160\,{{\it B_g}}^{2}\pi \,{R}^{4}-240\,{\it B_g}\,\pi \,{R}^{4}\rho_{{c}}-224\,{\it B_g}\,\pi \,{R}^{2}{r}^{2}\rho_{{0}}\nonumber \\
&+224\,{\it B_g}\,\pi \,{R}^{2}{r}^{2}\rho_{{c}}+80\,\pi \,{R}^{4}{\rho_{{c}}}^{2}+104\,\pi \,{R}^{2}{r}^{2}\rho_{{0}}\rho_{{c}}-104\,\pi \,{R}^{2}{r}^{2}{\rho_{{c}}}^{2}\nonumber \\
&+40\,\pi \,{r}^{4}{\rho_{{0}}}^{2}-80\,\pi \,{r}^{4}\rho_{{0}}\rho_{{c}}+40\,\pi \,{r}^{4}{\rho_{{c}}}^{2}+45\,{R}^{2}\alpha\,\rho_{{c}}+15\,{R}^{2}\rho_{{0}}\nonumber \\
& -15\,{R}^{2}\rho_{{c}}\big]\Bigg/\big[40\,\pi \,{R}^{2}\alpha\,{r}^{2}\rho_{{c}}-24\,\pi \,\alpha\,{r}^{4}
\rho_{{c}}+40\,\pi \,{R}^{2}{r}^{2}\rho_{{c}}+24\,\pi \,{r}^{4}\rho_{{0}}\nonumber \\
&-24\,\pi \,{r}^{4}\rho_{{c}}-15\,{R}^{2}\big].
\end{eqnarray}

\section{Physical features of the model}

From the above expression for the physical parameter $\Delta(r)$ [see FIG. 1], following the method of~\cite{Deb2016} and after using Eqs.~(\ref{eq16}) and (\ref{eq17}), we obtain the following equation in order to get the maximum anisotropy at the surface:
\begin{equation}
\begin{split}
{{\Delta'(r)|}_{r=R}}  = & -\frac{2}{9}\frac{1}{\left( 2\,M-R \right) ^{2} \left( \alpha+1 \right) {R}^{4}}\Bigg[-5376\,{B}^{2}M{\pi }^{2}{R}^{6}\alpha+1536\,{B}^{2}{\pi }^{2}{R}^{7}\alpha+3840\,{B}^{2}M{\pi }^{2}{R}^{6}-3072\,{B}^{2}{\pi }^{2}{R}^{7}+\\
& 3888\,B{M}^{2}\pi \,{R}^{3}\alpha-2544\,BM\pi \,{R}^{4}\alpha+336\,B \pi \,{R}^{5}\alpha-720\,B{M}^{2}\pi \,{R}^{3}+912\,BM\pi \,{R}^{4}-240\,B\pi \,{R}^{5} \\
& -540\,{M}^{3}\alpha+540\,{M}^{2}R\alpha-135\,M{R}^{2}\alpha-90\,{M}^{2}R+45\,M{R}^{2}\Bigg]=0. \label{eq25} 
\end{split}
\end{equation}

\begin{figure}[!htp]\centering
	\includegraphics[width=6cm]{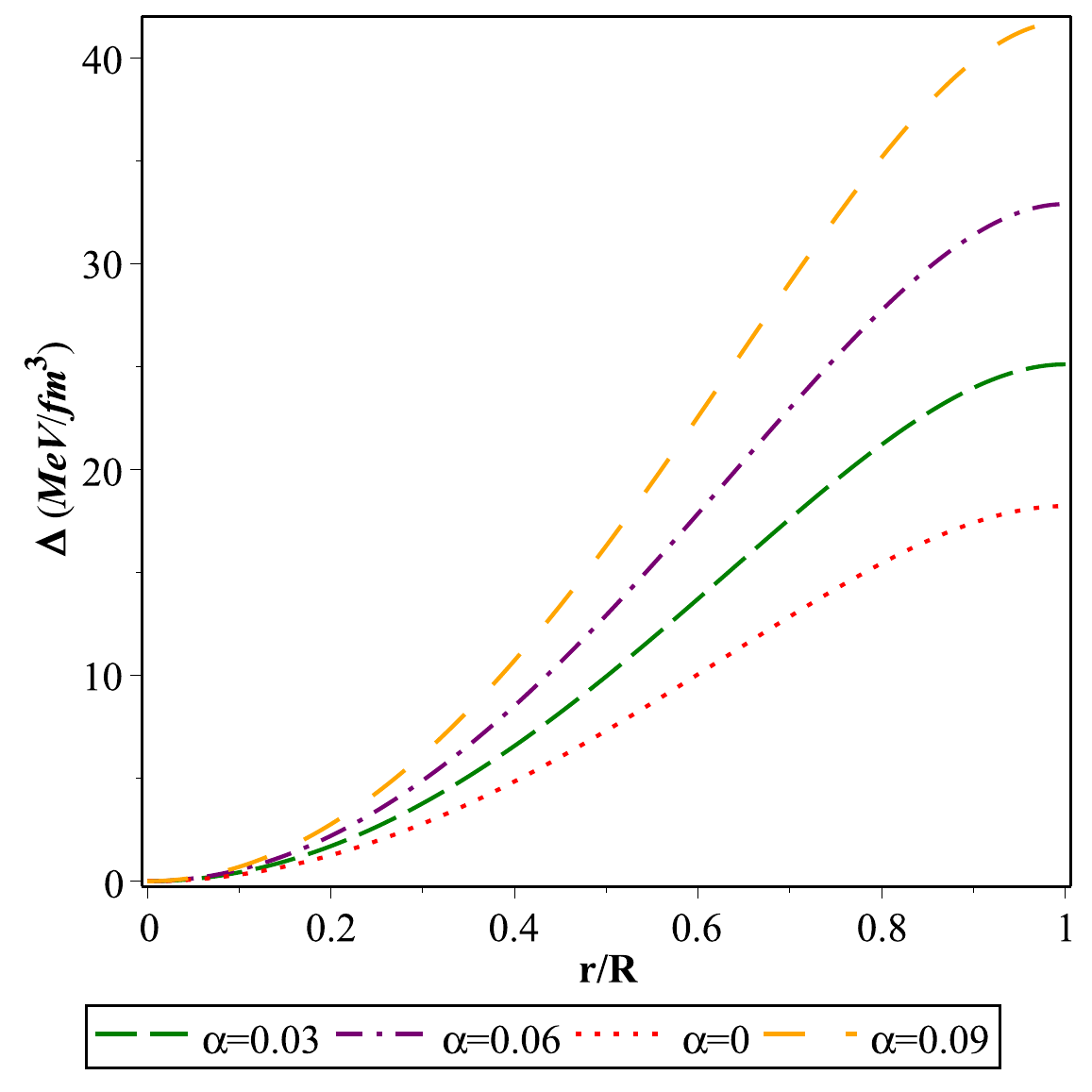}
	\caption{Variation of anisotropy $(\Delta)$ with the radial coordinate for different values of $\alpha$ for the strange star $SMC~X-4$, where $B_\text{g}$=83~$\text{MeV/(fm)}^3$, $R$=9.711~km and $M$= 1.29~$\text{M}_{\odot}$. }\label{Fig1}
\end{figure}


Solving Eq.~(\ref{eq25}), using different observational mass values for the various stars considered and with the choice of the parametric values of the bag constant as 83~$\text{MeV/(fm)}^3$~\cite{Rahaman2014} and $\alpha$, we get different values for the radius $R$ of the star. We choose only that value of $R$ which is physically valid and consistent with the Buchdahl condition~\cite{Buchdahl1959} and find that the anisotropy is maximum at the surface of the star. It is found that the central pressure (${p_{r}}^{eff} = {p_{t}}^{eff}$) is $3.691 \times 10^{34}~\text{dyne/cm}^2$ for $SMC~X-1$ due to $B_g=83~\text{MeV/fm}^3$ and $\alpha=0.06$. 

\begin{figure}[!htp]\centering
\includegraphics[width=10cm]{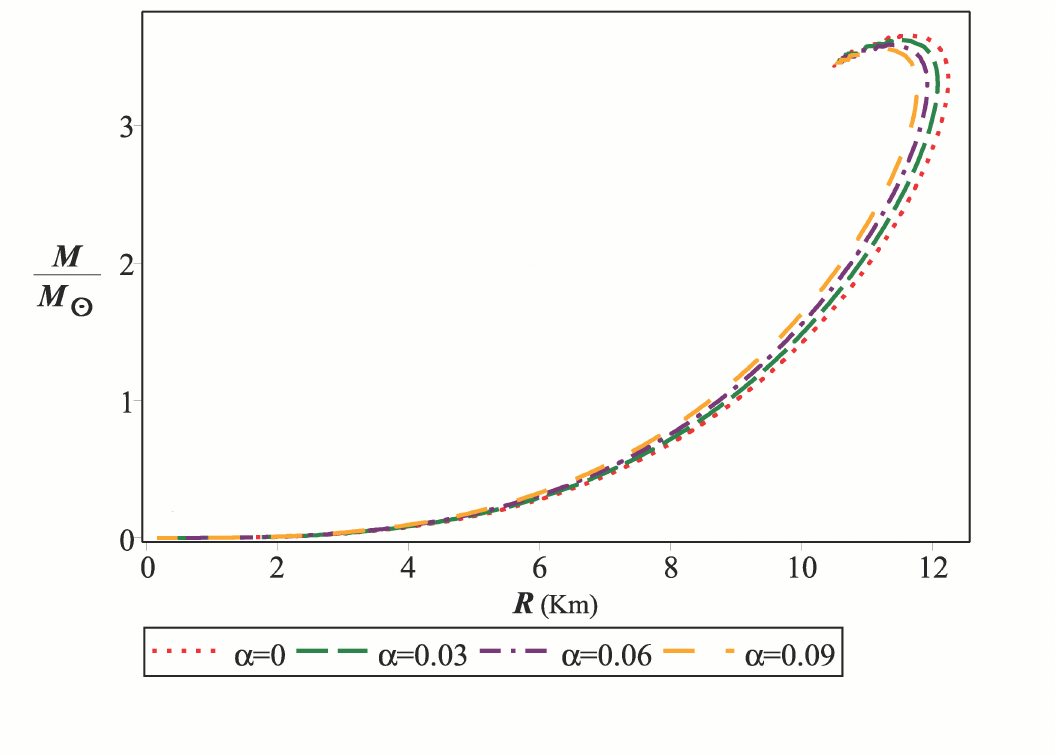}
\caption{Effective mass-radius $(M-R)$ curve for the different strange stars having  $B_\text{g}$=83~$\text{MeV/(fm)}^3$. The stars represented by the $M-R$ curve after it passes the maximum mass point are not stable. }\label{Fig2}
\end{figure}

According to Buchdahl~\cite{Buchdahl1959} the maximum allowed mass-radius ratio for a static spherically symmetric compact star is ${2\,M/R}\leq{8/9}$. Later Mak and Harko~\cite{Mak2003} came up with a more generalized expression for the same mass-radius ratio.
Now in our model the effective gravitational mass which is defined by Eq.~(\ref{eq16}) is given as $M=\frac{4}{15}\pi \,R \left( 2\,{R}^{2}\alpha\,\rho_{{c}}+3\,{R}^{2}\rho_{{0}}+2\,{R}^{2}\rho_{{c}} \right)$. The variation of the effective mass with the radius of the star is shown in FIG.~\ref{Fig2}.

Now the maximum value of the effective mass ${M}_\text{max}$  of the star corresponding to ${{{\rho}_{c}}|}_{{M}_\text{max}}$ can be derived using the equation $\frac{dM}{d{\rho}_{c}}=0$. Similarly, using the equation $\frac{dR}{d{{\rho}_{c}}}=0$ we can derive the maximum radius $R_\text{max}$ for ${{{\rho}_{c}}|}_{{R}_\text{max}}$. Due to $B_g=83~\text{MeV/fm}^3$ and $\alpha=0.06$ the maximum effective mass and the corresponding radius are $3.582~\text{M}_{\odot}$ and $11.301~\text{km}$, respectively.

To derive the entropy and the temperature of the stellar model, we are turning to the Gibbs relation, ${p^{eff}}+{{\rho}^{eff}}=s\,T+n\,\mu$, where $s(r)$ is the local entropy density, $T(r)$ is the local temperature, $\mu$ is the chemical potential, and $n$ is the number density of the matter distribution inside the ultra-dense star. Let us assume for the sake of simplicity, that the matter distribution inside the stellar configuration is isotropic in nature while the value of $\mu$ is negligible. Hence the Gibbs relation becomes
\begin{equation}
p^{eff}+{\rho}^{eff}=s\,T. \label{32}
\end{equation} 

Now using the first and the second laws of thermodynamics along with the first {\it ansatz}  we have the following relation: 
\begin{equation}
ds=\frac{V}{T}\,d{{\rho}^{eff}}+\frac{4}{3}\,\frac{({{\rho}^{eff}}-4\,{B_{g}})}{T}\,dV,\label{33}
\end{equation}
where $V$ is the volume of the stellar configuration. Since $S=S(\rho,V)$ and $dS$ is a total differential, one may find from Eq. (\ref{33})
\begin{equation}
{{\rho}^{eff}}=\beta\,{T}^{4}+B_{g},\label{34}
\end{equation}
where $\beta$ is the integration constant and $\sigma=\frac{1}{4}\,\beta$. Here $\sigma$ represents the famous Stefan-Boltzmann constant.

\begin{figure}[!htp]\centering
\includegraphics[width=6cm]{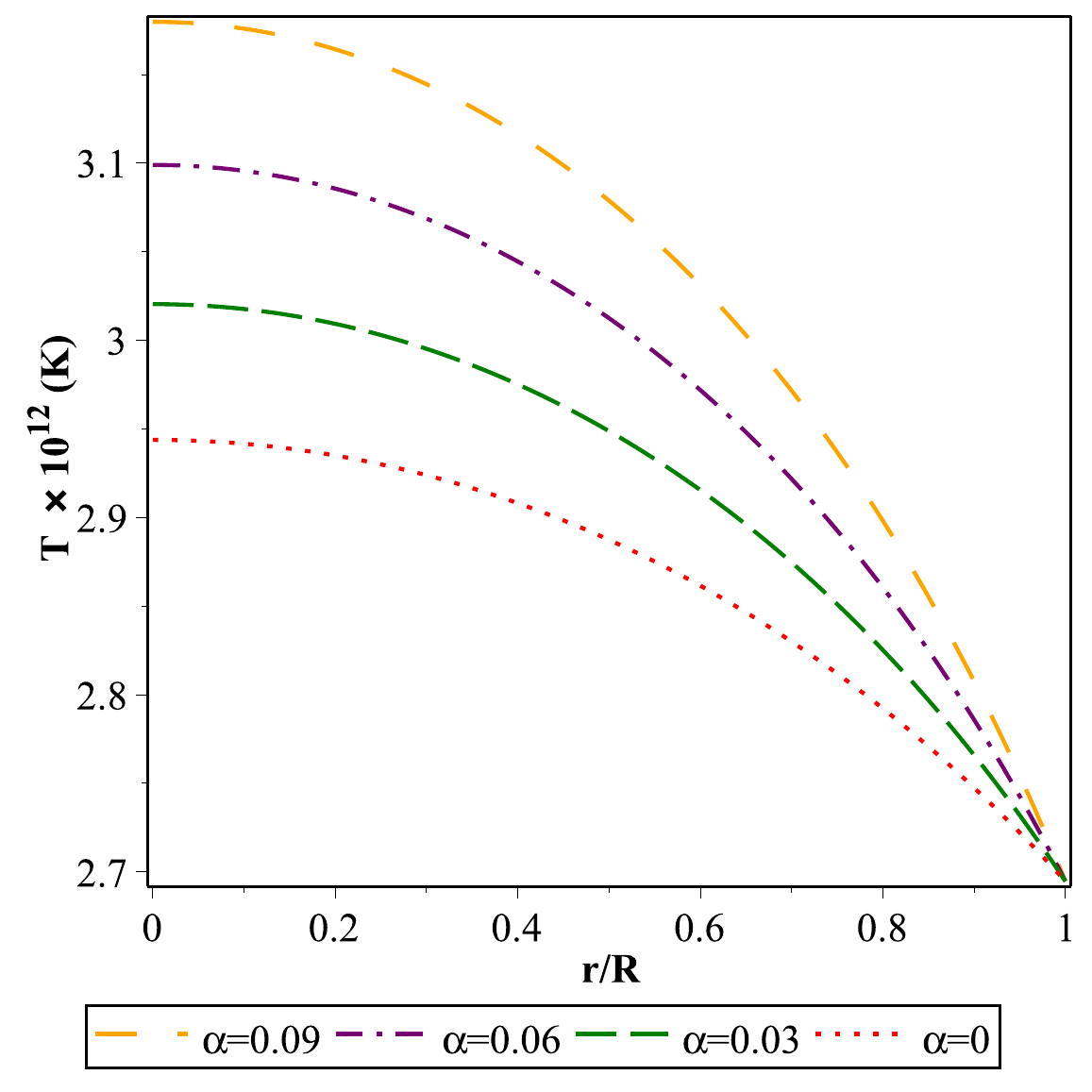}
\caption{The variation of temperature ($T$) with the radial coordinate for the  strange star $SMC~X-4$, where ${B_\text{g}}$=83~$\text{MeV/(fm)}^3$, $R$=9.711~km and $M$=1.29~~$\text{M}_{\odot}$.}\label{Fig3}
\end{figure}

Hence, after some algebra one can find the entropy density as 
\begin{equation}
s=\frac{4}{3}\,\beta\,{T}^{4}, \label{35}
\end{equation}
which provides a basic idea about the total entropy of the compact stellar system.

The variation of temperature in the interior region of the different strange star candidates is shown in FIG. \ref{Fig3}. From this figure, we see that the temperature in the central region is maximum while it decreases with the radial coordinate and becomes minimum at the surface, which is physically acceptable. For the stellar configurations, we find that the temperature is higher than the Fermi melting point $(0.5-1.2\times{{10}^{12}}$~K) for quarks. Hence all the quark matter in the ultra dense compact star remains in the form of quark-gluon plasma.

\begin{table*}
  \centering
    \caption{Numerical values of physical parameters for the different strange star candidates for $\alpha=0.06$ and $B=83~ MeV/{{fm}^3}$~\cite{Rahaman2014} } \label{Table 1}
   \scalebox{1.0}{
\begin{tabular}{ ccccccccccccccccccccccccccc}
 \hline Strange & Observed  & Predicted & \hspace{0.5cm}${\rho}^{\textit{{eff}}}_c$  & \hspace{0.3cm}${p^{\textit{{eff}}}_c}$ & \hspace{-0.1cm}$\frac{2M}{R}$ & $Z$ & \hspace{0.7cm}$T_c$  \\
        Stars  &  Mass (km) & Radius (km) & (${gm/cm}^3$) & \hspace{-0.2cm}(${dyne/cm}^2$)& & & \hspace{0.5cm}($K$) \\  
\hline $PSR~J~1614-~2230$ &~~ $1.97 \pm 0.04$ &~~  $10.703^{+0.060}_{-0.061}$  & $1.018 \times {10}^{15} $ & $ 5.852 \times {10}^{34} $ &  $0.543$ &  $0.479$ & $3.188\times {10}^{12}$
\\ \hline $Vela~X-1$  &~~ $1.77 \pm 0.08$ &~~    $10.384^{+0.132}_{-0.139}$  & $ 9.868 \times 10^{14} $ & $5.139 \times 10^{34}$ & $0.503$ & $0.418$ & $3.159 \times {10}^{12}$\\ 
\hline $PSR~J~1903+~327$ &~~ $1.667 \pm 0.021$   &~~  $10.205^{+0.037}_{-0.038}$ & $9.723 \times 10^{14}$ & $4.802 \times 10^{34}$ &  $0.482$ & $0.389$ & $3.145 \times {10}^{12}$\\ 
\hline $Cen~X-3$ &~~ $1.49 \pm 0.08$   &~~  $9.871^{+0.155}_{-0.162}$ & $9.493 \times 10^{14}$ & $4.270 \times 10^{34}$ &  $0.445$ & $0.342$ & $3.124 \times {10}^{12}$\\ 
\hline $SMC~X-1$ &~~ $1.29 \pm 0.05$   &~~  $9.451^{+0.110}_{-0.114}$ & $9.243 \times 10^{14}$ & $3.691 \times 10^{34}$ & $0.403$ & $0.294$ & $3.099 \times {10}^{12}$\\ 
\hline
\end{tabular}  }
  \end{table*}

For the anisotropic static stellar configuration, though radial pressure vanishes at the surface, the tangential pressure does not. However, as the radial pressure is continuous at the boundary, we already satisfy Synge's junction condition~\cite{Synge1952} in the case of static spherical symmetry. In the boundary, the interior solution and the exterior Schwarzschild solution should match in order to satisfy the fundamental junction condition. The metric coefficients are continuous at the surface $S$ where $r=R=$ constant. The second fundamental form is also continuous on the boundary surface. Now the intrinsic stress-energy tensor $S^{i}_j=\text{diag}(-\sigma,\mathcal{P})$ at the boundary surface $S$ (where $r = R$) can be defined as the surface stresses, i.e the surface energy $\sigma$ and the surface tangential pressures $ p_\theta =p_\phi \equiv \mathcal{P}$ which in the present situation are given as $\sigma = 0$ and $\mathcal{P} = 0$. 
Thus the complete spacetime is given by our interior metric and the exterior Schwarzschild metric, which are matched smoothly on the boundary surface $S$.

\section{Discussion and conclusions}

Under the proposed model we have presented a data set for the physical parameters of some strange star candidates in TABLE I.

Let us highlight the major results of the proposed model: (i) the quark matter is converting into dark energy under a certain critical condition; (ii) with the presence of dark energy inside a strange star, the latter behaves like a dark energy star; (iii) the high temperature distribution ($>$Fermi melting point for quarks) inside the star confirms the presence of quark matter in the form of quark-gluon plasma; (iv) all the physical and structural features of the proposed ultra-dense strange star model match well with a dark energy star as suggested by Chapline~\cite{Chapline2004}; and (v) some of the physical tests, such as the energy conditions, the TOV equations and the sound speed constraint are found to be satisfied in the presented model, and thus the model has a stable configuration in all respects. According to Herrera~\cite{Herrera1992} and Andr{\'e}asson~\cite{Andreasson2009} in order to form a physically acceptable matter distribution, the quark matter also has to respect the condition  $0 \leq{v_\text{sqr}}^2 \leq 1$, where $v_{sqr}$ represents the radial sound speed of the quark matter. This leads to the result that the acceptable value of $\alpha$ lies in the range $0 \leq \alpha \leq 0.11$.

An obvious query regarding the present investigation may be as follows: are there any current or future missions that could provide us with some data that would allow us to probe the inner structure of relativistic compact stars? The recent observation of two gravitational wave (GW) events, GW170817~\cite{Abbott2017} and GW190425~\cite{Abbott2020} put a well-defined constraint on the EOS of the neutron stars as $\tilde{\Lambda}\lesssim 800$, where $\tilde{\Lambda}$ is the effective tidal deformability of the binary neutron star system, which leads to ruling out of the EOS that supports radii above 13~km for the $1.4~M_\odot$ neutron stars. In fact, observational data from GW favoured soft EOS and ruled out stiff EOS, such as ms1b, h4, etc. On the other hand, the observation data from the Neutron Star Interior Composition Explorer (NICER)~\cite{Miller2019} sets stringent constraints on the radius of the neutron stars, which supports the presence of stiffer EOS. Interestingly, this dichotomy of the results from GW observations and NICER data become helpful to understand the true nature of EOS of the compact stars and their interior structure. Recent observational constraints suggest that primordial black holes (PBHs)with mass scales $ \sim 10^{-12} M_{\odot} $ can account for the vast majority of dark matter (DM) in the Universe. We need an increase in primordial scalar curvature perturbations to the order of $ \mathcal{O}(10^2)$ at the scale $ k \sim 10^{12} Mpc^{-1} $ to produce these PBHs~\cite{Solbi}. Based on the analysis of years of data from the Kepler satellite, searching for short-duration bumps induced by gravitational micro-lensing, unique limits on the acceptable masses of a DM halo composed of PBHs and/or any other massive compact halo object have been imposed~\cite{Griest}. The masses range from $ 2 \times 10^{-9} M_{\odot}$ to $ 10^{-7} M_{\odot}$. Prospective analyses of the whole Kepler data set should find PBH DM or rule out some of this spectrum, and space missions like WFIRST Wide-Field Infrared Survey Telescope (WFIRST) have the power to reach an order of magnitude more.

Let us comment on some shortcomings of the proposed model, which do exist as follows: \\ (i) we proposed that quark matter is converting into dark energy, but we cannot predict under which condition this is actually happening, and (ii) in the range $0\leq \alpha \leq 0.11$, we consider $\alpha$ as a constant parametric term. Therefore it is difficult to predict through this model whether the dark energy in the stellar configuration remains constant or it varies with time within the provided range of $\alpha$. In connection to this comment, one can specifically note that the metric coefficients are considered independent of time in the present model. However, all these issues can be considered in a future investigation.\\

\section*{Data Availability Statement}
No new data were generated or analysed in support of this research.

\section*{Conflict of Interest Statement}
There is no conflict of interest in connection to the present work.

\section*{Acknowledgments} 
We all are grateful to Professor A. DeBenedictis, Simon Fraser University, for several valuable suggestions. FR and SR are thankful to the Inter-University Centre for Astronomy and Astrophysics (IUCAA), Pune, India for providing Visiting Associateship under which a part of this work was carried out. Research of DD is funded by the C.V. Raman Postdoctoral Fellowship (Reg. No. R(IA)CVR-PDF/2020/222) from the Department of Physics, Indian Institute of Science. SRC would like to thank the Southern Federal University (SFedU) for financial support (Grant No. P-VnGr/21-05-IF). The presentation of the manuscript in `Can strange stars mimic dark energy stars?' with arXiv: 1611.02253 is a part of the present article with revised authorships and version~\cite{Gram}.


\begin{thebibliography}{99}

\bibitem{Bhattacharyya1999} A.~Bhattacharyya, {\sl et al.}, Nucl. Phys. A \textbf{661}, 629 (1999).

\bibitem{Bhattacharyya2000} A.~Bhattacharyya, {\sl et al.}, Phys. Rev. D \textbf{61}, 083509 (2000).

\bibitem{Bhattacharyya2003} A.~Bhattacharyya, {\sl et al.}, Pramana: J. Phys. \textbf{60}, 909 (2003).

\bibitem{Witten1984} E.~Witten, Phys. Rev. D \textbf{30}, 272 (1984).

\bibitem{Applegate1985} A.~Applegate and C.J.~Hogan, Phys. Rev. D \textbf{31}, 3037, (1985).

\bibitem{Farhi1984} E.~Farhi and R.L.~Jaffe, Phys. Rev. D \textbf{30}, 2379 (1984).

\bibitem{Chandra2000} D.~Chandra and A.~Goyal, Phys. Rev. D \textbf{62}, 063505 (2000).

\bibitem{Ghosh2008} S.~Ghosh, Astrophysics of Strange Matter, Plenary Talk at 2008 Quark Matter, Jaipur, India;
arXiv:astro-ph/0807.0684.

\bibitem{Perez-Garcia2010} M.A.~Perez-Garcia, J.~Silk and J.R.~Stone, Phys. Rev. Lett. \textbf{105}, 141101 (2010).

\bibitem{Drake2002} J.J.~Drake, {\sl et al.}, Astrophys. J. \textbf{572}, 996 (2002).

\bibitem{Madsen1999} J.~Madsen, Lect. Notes Phys. \textbf{516}, 162 (1999).

\bibitem{Farook2012} F.~Rahaman, {\sl et al.}, Phys. Lett. B \textbf{714}, 131 (2012).

\bibitem{Brilenkov2013} M.~Brilenkov, {\sl et al.}, JCAP \textbf{08}, 002 (2013). 

\bibitem{Adamovich1991} M.I. Adamovich, {\sl et al.}, Phys. Lett. B \textbf{263}, 539 (1991).

\bibitem{Chapline2004} G.~Chapline: Proc. of the 22nd Texas Symposium on Relativistic Astrophysics, 2004; arXiv: astro-ph/0503200.

\bibitem{Chapline2001} G.~Chapline, E.~Hohlfeld, R.B.~Laughlin and D.~Santiago, Philos. Mag. B, \textbf{81}, 235 (2001).

\bibitem{Barbierii2004} J.~Barbierii and G.~Chapline, Phys Lett. B, \textbf{590}, 8 (2004).

\bibitem{Deb2016} D.~Deb, {\sl et al.}, Ann. Phys. \textbf{387}, 239 (2017).

\bibitem{Davies1984} C.W.~Davies, Phys. Rev. D \textbf{30}, 737 (1984).

\bibitem{Blome1984} J.J.~Blome and W.~Priester, Naturwiss. \textbf{71}, 528 (1984).

\bibitem{Hogan1984} C.~Hogan, Nat. \textbf{310}, 365 (1984).

\bibitem{Kaiser1984} N.~Kaiser and A.~Stebbins, Nat. \textbf{310}, 391 (1984).

\bibitem{Harko2002} M.K.~Mak and T.~Harko, Chine. J. Astron. Astrophys. \textbf{2}, 248 (2002).

\bibitem{Rahaman2014} F.~Rahaman, {\sl et al.}, Eur. Phys. J. C {\bf 74}, 3126 (2014).

\bibitem{Buchdahl1959} H.A.~Buchdahl, Phys. Rev. D \textbf{116}, 1027 (1959).

\bibitem{Mak2003} M.K.~Mak and T.~Harko, Proc. R. Soc. A \textbf{459}, 393 (2003).

\bibitem{Synge1952} S.~ O'Brien S. and J.L.~Synge, Commun. Dublin Inst. Adv. Stud. A. {\bf 9} (1952).

\bibitem{Herrera1992} L.~Herrera, Phys. Lett. A {\bf 165}, 206 (1992).

\bibitem{Andreasson2009} H.~Andr{\'e}asson, Commun. Math. Phys. {\bf 288}, 715 (2009).

\bibitem{Abbott2017} B.P.~Abbott, R.~Abbott, T.D.~Abbott et al., Phys. Rev. Lett. \textbf{119}, 161101 (2017).

\bibitem{Abbott2020} B.P.~Abbott, R.~Abbott, T.D.~Abbott et al., Astrophys. J. Lett. 892, L3 (2020).

\bibitem{Miller2019} M.C.~Miller, F.K.~Lamb, A.J.~Dittmann et al., Astrophys. J. Lett. \textbf{887}, L24 (2019).

\bibitem{Rahaman2014} F.~Rahaman, K.~Chakraborty, P.K.F.~Kuhfittig, G.C.~Shit and M.~Rahman, Eur. Phys. J. C  \textbf{74}, 3126 (2014).

\bibitem{Solbi} M.M.~Solbi and K.~Karami, JCAP \textbf{08}, 056 (2021).

\bibitem{Griest} K.~Griest, A.M.~Cieplak and M.J.~Lehner, Phys. Rev. Lett. \textbf{111}, 181302 (2013).

\bibitem{Gram} T.~Grammenos, F.~Rahaman, S.~Ray, D.~Deb and S.~Roy Chowdhury, Can strange stars mimic dark energy stars? https://arxiv.org/pdf/1611.02253.

\end{thebibliography}
\end{document}